\documentclass[twocolumn]{aastex62}
\usepackage[fleqn]{amsmath}
\usepackage{hyperref}
\newcommand{\acsi}{\it F814W}
\newcommand{\acsv}{\it F555W}
\newcommand{\acsr}{\it F606W}

\shorttitle{TRGB in the LMC} 
\shortauthors{Yuan et al.}

\begin{document}

\title{Consistent Calibration of the Tip of the Red Giant Branch in the Large Magellanic Cloud on the {\it Hubble Space Telescope} Photometric System and a Re-determination of the Hubble Constant}

\author[0000-0001-9420-6525]{Wenlong Yuan}
\affiliation{Department of Physics \& Astronomy, Johns Hopkins University, Baltimore, MD, USA}
\author{Adam G.~Riess}
\affiliation{Department of Physics \& Astronomy, Johns Hopkins University, Baltimore, MD, USA}
\affiliation{Space Telescope Science Institute, Baltimore, MD, USA}
\author[0000-0002-1775-4859]{Lucas M.~Macri}
\affiliation{George P.~and Cynthia W.~Mitchell Institute for Fundamental Physics \& Astronomy,\\Department of Physics \& Astronomy, Texas A\&M University, College Station, TX, USA}
\author{Stefano Casertano}
\affiliation{Space Telescope Science Institute, Baltimore, MD, USA}
\author{Daniel M. Scolnic}
\affiliation{Duke University, Department of Physics, Raleigh, NC, USA}

\email{wyuan10@jhu.edu}

\begin{abstract}

 We present a calibration of the Tip of the Red Giant Branch (TRGB) in the Large Magellanic Cloud (LMC) on the $HST$/ACS {\acsi} system. We use archival $HST$ observations to derive blending corrections and photometric transformations for two ground-based wide-area imaging surveys of the Magellanic Clouds. We show that these surveys are biased bright by up to $\sim$0.1 mag in the optical due to blending, and that the bias is a function of local stellar density. We correct the LMC TRGB magnitudes from \citet{2017ApJ...835...28J} and use the geometric distance from \citet{2019Natur.567..200P} to obtain an absolute TRGB magnitude of $M_{F814W}=-3.97\pm0.046$~mag. Applying this calibration to the TRGB magnitudes from \citet{2019arXiv190705922F} in SN Ia hosts yields a value for the Hubble constant of $H_0=72.4\pm2.0$ km s$^{-1}$ Mpc$^{-1}$ for their TRGB$+$SNe Ia distance ladder. The difference in the TRGB calibration and the value of $H_0$ derived here and by \citet{2019arXiv190705922F} primarily results from their overestimate of the LMC extinction, caused by inconsistencies in their different sources of TRGB photometry for the Magellanic Clouds. Using the same source of photometry (OGLE) for both Clouds and applying the aforementioned corrections yields a value for the LMC $I$-band TRGB extinction that is lower by 0.06 mag, consistent with independent OGLE reddening maps used by us and by \citet{2017ApJ...835...28J} to calibrate TRGB and determine $H_0$. \\ 
  
\end{abstract}

\section{Introduction}

The Large Magellanic Cloud (LMC) provides a cornerstone in the efforts to calibrate the luminosities of standard candles. It is near enough for its distance to be measured geometrically \citep{2019Natur.567..200P} and its geometry is well-understood so that imaging of its stars is readily converted into useful estimates of their luminosities.   Wide-area surveys towards the LMC, such as MACHO \citep{2000ApJ...542..281A}, OGLE \citep{1997AcA....47..319U}, the LMC Near-Infrared Synoptic Survey \citep{2015AJ....149..117M} and the VMC survey \citep{2011A&A...527A.116C} provide high-quality optical to near-infrared photometric measurements of millions of LMC stars, offering the opportunity to calibrate distance indicators such as Cepheid variables \citep{2008AcA....58..163S, 2015AJ....149..117M}, the Tip of the Red Giant Branch \citep[TRGB;][]{2017ApJ...835...28J,2018ApJ...858...12H,2018arXiv181202581G}, Miras \citep{2009AcA....59..239S,2017AJ....154..149Y}, RR Lyraes \citep{2009AcA....59....1S}, and Population II pulsators \citep{2017AJ....153..154B}. As a result, the LMC has been one of the most critical anchors for calibrating extragalactic distance indicators and measuring the Hubble constant, $H_0$ \citep{2001ApJ...553...47F, 2006ApJ...653..843S, 2011ApJ...730..119R, 2016ApJ...826...56R, 2017ApJ...835...28J, 2019ApJ...876...85R, 2019arXiv190705922F}.

Given a geometric distance measurement to the LMC now approaching a total uncertainty of only one percent \citep{2019Natur.567..200P}, efforts to extend flux and distance measurements based on these standard candles but using different telescopes and cameras are now limited by {\it cross-instrument} photometric uncertainties. For example, \citet{2019ApJ...876...85R} found a 0.02-0.04~mag offset between {\it HST} and ground measurements for 70 LMC Cepheids when directly imaging these stars from both platforms, with the value of the offset depending on the bandpass and the ground survey. Differences will naturally arise due to inconsistencies between photometric systems, differences in system bandpasses and differences in resolution. Importantly, such offsets will be system-dependent and thus must be considered anew for each ground-based survey, the relevant spectral energy distribution, source brightness and density of contaminating stars.  As we will show, these offsets are often larger than the current precision of the measurement of $H_0$ ($\sim$ 0.04 mag) and need to be addressed for other distance indicators besides Cepheids.  To make optimal use of the exquisite LMC geometric distance estimate to anchor a long-range distance ladder, we need to either directly rely on space observations or robustly measure the relevant cross-instrument offset. 

The Advanced Camera for Surveys (ACS) onboard {\it HST} has been extensively used to acquire extragalactic TRGB measurements \citep[e.g.][]{2006AJ....132.2729M,2008ApJ...689..721M,2017ApJ...835...28J,2017ApJ...845..146H}. Most of these optical TRGB studies made use of the {\it HST} equivalent of the $I$-band filter, {\acsi}, together with a bluer filter (usually {\acsv} or {\acsr}) to constrain the RGB sample and obtain distances based on color and the {\acsi} TRGB magnitude.    As the LMC TRGB stars are very bright for {\it HST} and widely spread across many square degrees, it is regrettably not feasible to directly measure the TRGB of the LMC in the {\it HST} photometric system.  Thus the lack of a wide and shallow {\it HST} survey of the LMC precludes a direct {\acsi} TRGB zero point calibration against geometric distance. Deeper {\it HST} imaging is not useful to compare to the ground due to the lack of overlapping fluxes.  For stars near the TRGB ($I \sim 14.5$ mag), ACS {\acsi} saturates in only 20 seconds and it is important to measure even brighter stars to detect the ``edge'' of the tip, limiting the past {\it HST} imaging relevant for a cross-calibration to a small set.  However, there is one program (GO 9891, PI Gerard Gilmore) that observed a dozen LMC fields with exposure times of 20 to 120 seconds with ACS {\acsi} and {\acsv} which overlap ground programs that fully cover the LMC, allowing a more direct calibration the TRGB in the LMC for the same instrument and filter used in extragalactic work.

\begin{figure*}
\epsscale{1.3}
\includegraphics[width=\textwidth]{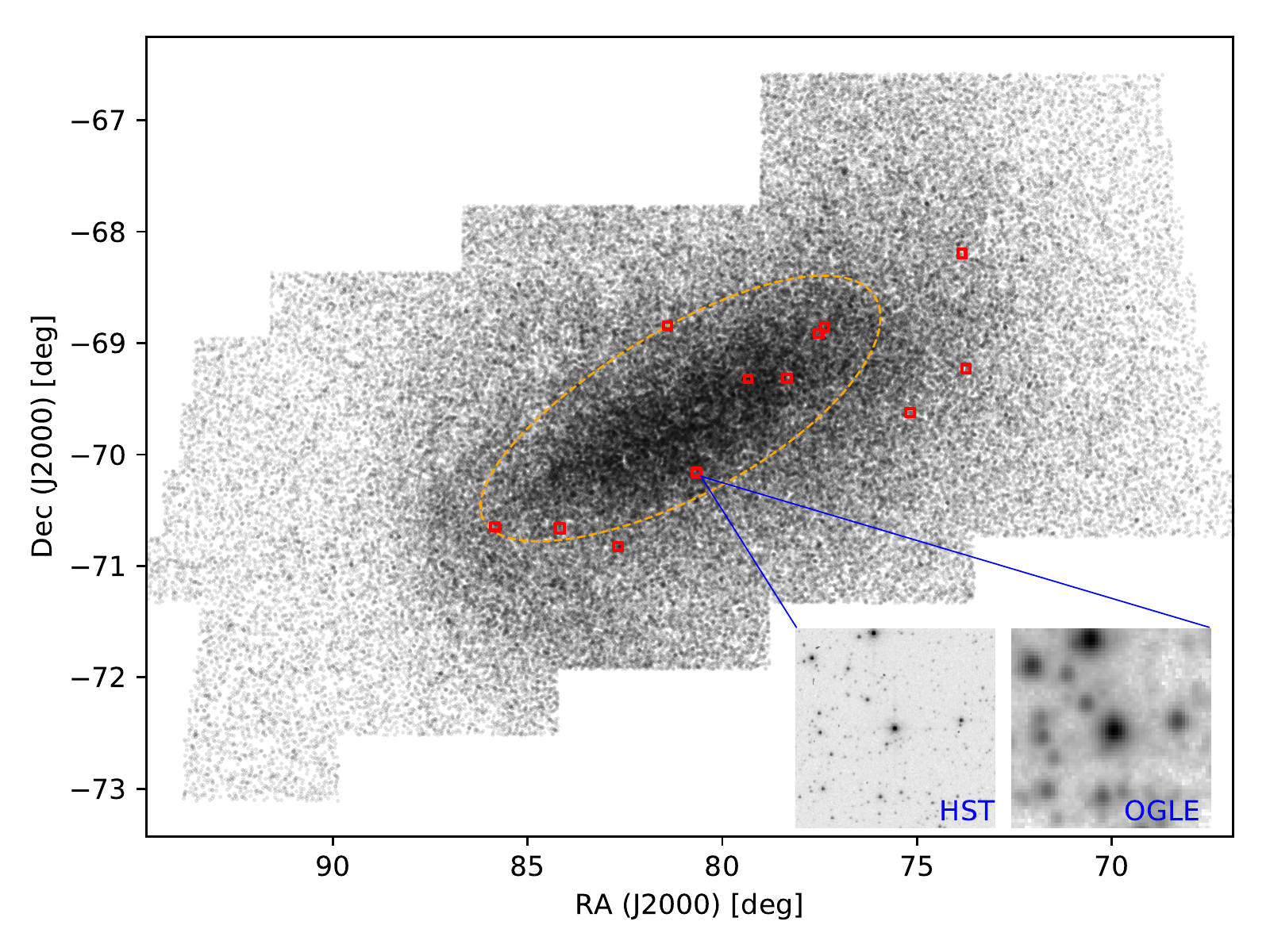}
\caption{Locations of 12 {\it HST} fields used in this study (red squares). The size of the squares is in scale of the {\it HST} field of view. The lower right image cuts show the comparison of resolving power between the ACS {\acsi} and OGLE $I$-band reference image which was observed at top seeing condition. The yellow dashed ellipse defined the ``bar region'' adopted in this work.\label{fig_lmc}}
\end{figure*}

In this work, we address the ground-to-{\it HST} zeropoint offset of the TRGB method for two commonly-used ground-based surveys and obtain the transformed TRGB magnitude in the {\it HST} ACS {\acsi} system. The ground surveys we studied are the third phase of the Optical Gravitational Lensing Experiment \citep[OGLE-III;][]{2008AcA....58...89U} and the Magellanic Clouds Photometric Survey \citep[MCPS;][]{2004AJ....128.1606Z}. The plate scales of these two surveys are 0.27 and 0.7{\arcsec}/pix, with typical seeing of 1.2 and 1.5{\arcsec}, respectively. In crowded regions in the LMC, such as its bar, we expect unresolved flux could contaminate the photometric measurements of these surveys. In comparison, the resolving power of ACS or WFC3 on {\it HST} is much higher, with pixel scales of 0.04-0.05\arcsec and near diffraction-limited resolution of $\sim 0.1$\arcsec, allowing LMC stars to be easily resolved and accurately photometered.

Figure~\ref{fig_lmc} shows a comparison of {\it HST}/ACS and OGLE-III images of the same region of the LMC.  Although images from the MCPS survey are not available, the resolution is significantly worse than the displayed OGLE-III reference image which is a stack of $\sim$10 observations with seeing $<1$\arcsec.  To determine the LMC TRGB in the {\it HST}/ACS system, we studied the blending effect by comparing the ground and archival {\it HST} observations for stars fainter than and up to the edge of the TRGB ($14.5 \lesssim I \lesssim 17$~mag). We used the local stellar number density and the magnitude of each star to characterize its blending, and measured the ground-to-{\it HST} magnitude offset as a function of stellar density, magnitude and color. We find a large ($\sim0.1$~mag) offset in the TRGB magnitude determined from MCPS data and a much smaller but still non-negligible offset for OGLE III.  We demonstrate that the offset is a statistical blending effect due to the limited resolving power of ground instruments, as well as different filter responses for red stars.

The rest of the paper is organized as follows: \S2 describes the data sources and reduction of the {\it HST} data; \S3 gives the methodology of ground-to-{\it HST} magnitude transformation; \S4 presents the mean ground-to-{\it HST} offset of various regions across the LMC and derives the LMC TRGB value in the ACS {\acsi} system. We also discuss in that section the impact of the offset on the estimate of extinction towards the LMC and re-determine the TRGB-based estimate of $H_0$.

\section{Data and photometry}

In order to obtain the ground-to-{\it HST} correction for the TRGB in the LMC, we analyzed two commonly-used ground-based photometric catalogs: OGLE-III \citep{2008AcA....58...89U} and the MCPS \citep{2004AJ....128.1606Z}, as well as archival {\it HST} observations of 12 LMC fields that are covered by these ground-based surveys.

We retrieved the photometry catalog of OGLE-III for the LMC, which is based on repeated $VI$ observations covering 40 square degrees. Their catalog includes astrometry and photometry of $\sim 3.5\times 10^7$ sources. We refer interested readers to \citet{2008AcA....58...69U} for a detailed description of the photometry and calibration. In this study we made a few cuts to select sources of interest. We firstly selected a large subsample (hereafter the ``density sample'') by excluding sources with $I>18$~mag and merging duplicate sources in overlapping subfields within a radius of one pixel ($0.27\arcsec$). This sample was used to derive the local stellar number density. We further selected a ``TRGB calibration sample'' with $I<17$~mag and $V\!-\!I > 0.5$~mag to exclude sources less relevant to the TRGB. The latter sample also excluded sources with large standard deviations in mean magnitude ($\sigma_I\,>\,0.2$~mag, $\sigma_V\,>\,0.4$~mag), which tend to be large-amplitude AGB variables. The density and calibration samples consist of $\sim 1.9\times 10^6$ and $\sim 6.5\times 10^5$ objects, respectively. We performed a similar selection for the MCPS and obtained a ``calibration sample'' of $\sim 8.2\times 10^5$ objects.

\begin{figure}
\plotone{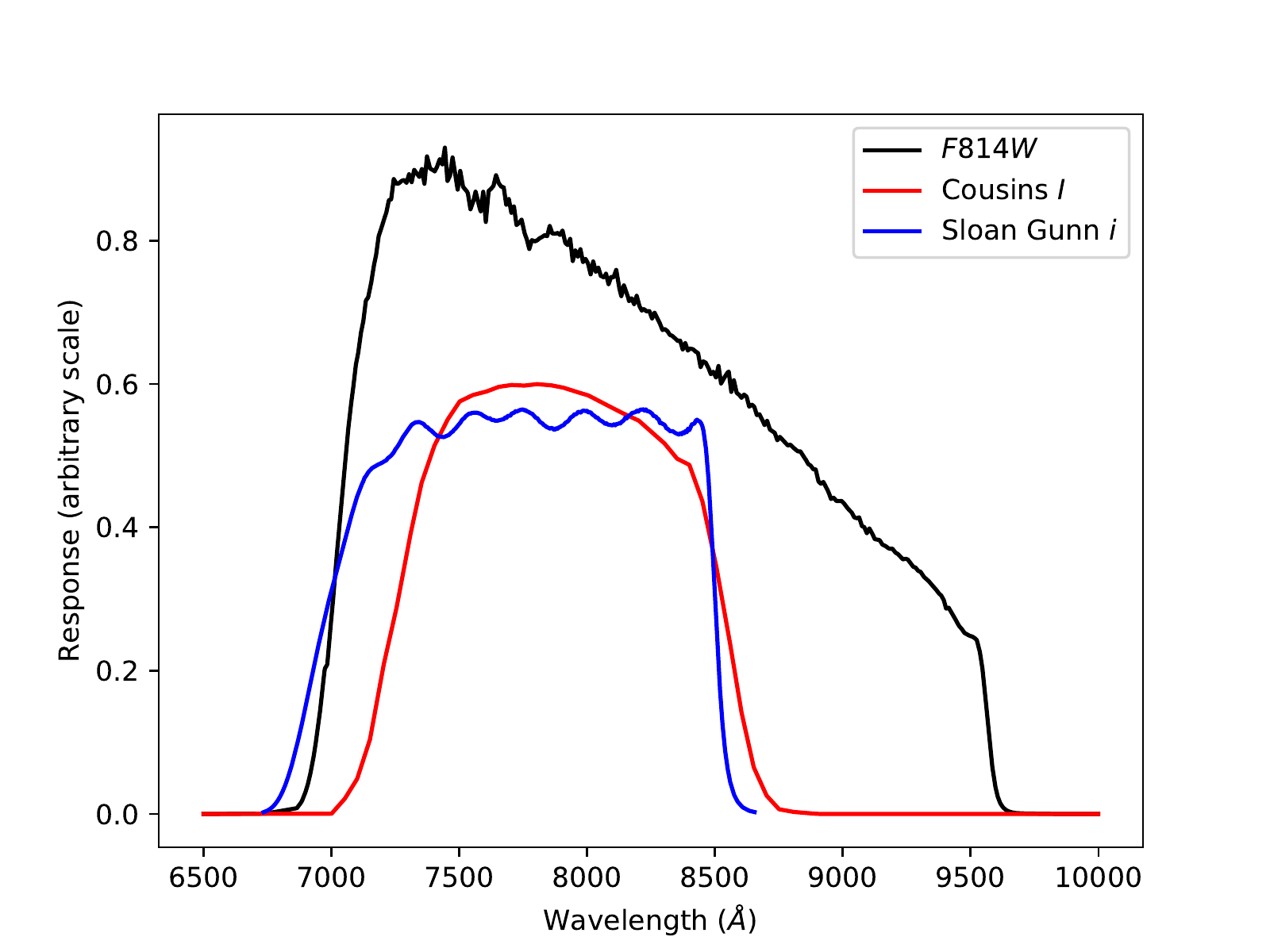}
\caption{Filter response curves for ACS {\acsi} (black), Cousins $I$ (red), and Sloan Gunn $i$ (blue), which are used by the {\it HST}, OGLE, and MCPS observations, respectively. The responses are set to arbitrary scale. \label{fig_filters}}
\end{figure}

There are two important differences between the MCPS and the OGLE data. Firstly, MCPS used a ``Sloan Gunn $I$'' filter, which is bluer than the Cousins $I$ (an example of this type of filter is shown in Figure~\ref{fig_filters}). Although the MCPS magnitudes were placed on the Johnson-Kron-Cousins system, Figure~1 of \citet{2002AJ....123..855Z} indicates large color residuals for red stars which are more relevant for this study. The trace of this filter has not been published and is not available, making it difficult to assess (Zaritzky 2019, private communication). Secondly, the pixel size of MCPS is larger than that of OGLE by a factor of 2.6 (an area ratio of 6.7), and its seeing was worse than that of OGLE, resulting in greater blending. These differences make the MCPS sample a less favorable source compared to OGLE for TRGB calibration. However, we present the MCPS 
ground-to-{\it HST} transformation because \citet[][hereafter F19]{2019arXiv190705922F} recently used MCPS photometry of the Small Magellanic Cloud (SMC) to determine the extinction of TRGB in the LMC. We will demonstrate in \S~\ref{sec_met} the significant blending effect of the MCPS data, and present in \S~\ref{sec_imp} that using MCPS in the SMC but OGLE in the LMC without accounting for their inconsistencies, as done by F19, caused a bias in their extinction corrected TRGB zeropoint.

\begin{deluxetable}{ccccrr}
\tabletypesize{\scriptsize}
\tablecaption{Summary of {\it HST} Observations\label{tbl_obs}}
\tablewidth{0pt}
\tablehead{
\colhead{Field} & \colhead{Date} & \colhead{RA} & \colhead{Dec} & \multicolumn{2}{c}{Exp.~time [sec]} \\[-6pt]
\colhead{(NGC)} &               & \multicolumn{2}{c}{(J2000) [deg]}& \acsv & \acsi}
\startdata
1755 &   2003-08-08 &     73.81569 &    -68.20538 &    50 &    40 \\
1756 &   2003-08-12 &     73.70745 &    -69.23729 &   170 &   120 \\
1801 &   2003-10-08 &     75.14557 &    -69.61352 &   115 &    90 \\
1854 &   2003-10-07 &     77.33527 &    -68.84784 &    50 &    40 \\
1858 &   2003-10-08 &     77.49211 &    -68.90496 &    20 &    20 \\
1872 &   2003-09-21 &     78.29894 &    -69.31275 &   115 &    90 \\
1903 &   2004-06-02 &     79.34387 &    -69.33769 &    50 &    40 \\
1943 &   2003-10-07 &     80.62299 &    -70.15447 &    50 &    40 \\
1953 &   2003-10-07 &     81.36727 &    -68.83779 &   115 &    90 \\
1983 &   2003-10-07 &     81.93727 &    -68.98526 &    20 &    20 \\
2010 &   2003-10-07 &     82.64549 &    -70.81896 &    20 &    20 \\
2056 &   2003-08-08 &     84.14355 &    -70.67142 &   170 &   120 \\
2107 &   2003-10-07 &     85.80286 &    -70.64053 &   170 &   120
\enddata
\end{deluxetable}

We used MAST to retrieve archival {\it HST} ACS {\acsi} and {\acsv} observations of 13 fields that are covered by the aforementioned ground surveys. These fields are centered on globular clusters NGC$\,$1755, 1756, 1801, 1854, 1858, 1872, 1903, 1943, 1953, 1983, 2010, 2056 and 2107, but the majority of each frame covers non-cluster stars. The observations are summarized in Table~\ref{tbl_obs} and their locations are shown in Figure~\ref{fig_lmc}. The images were processed with the standard {\it HST} calibration pipeline and calibrated for flat field, charge transfer efficiency, geometric distortion, pixel area, etc. We found the image quality of NGC$\,$1983 is noticeably worse than the rest of the fields and thus excluded it from the analysis. Since these fields are not very crowded at the resolution of {\it HST}, and because only bright sources were used for the analysis, we found that aperture photometry is robust and gives accurate measurements. We obtained 4-pixel aperture photometry using DAOPHOT \citep{1987PASP...99..191S}, followed by 10-pixel aperture corrections of $-0.111$~mag for {\acsi} and $-0.092$~mag for {\acsv} that are determined from isolated bright stars in these fields. We finally added the time-dependent Vega~mag zeropoint using the {\tt pysynphot 0.9.12} program which corrects the 10-pixel aperture instrumental magnitudes to infinite-aperture Vega magnitudes. In summary, the magnitude calibration is set by ({\acsi} as an example)
\begin{equation*}
  m = m_\mathrm{4pix} - 0.111 + \Delta m_\mathrm{inf} + \mathrm{ZP}_\mathrm{Vega}(\mathrm{MJD}),
\end{equation*}
where $\Delta m_\mathrm{inf}$ ($-0.0965$ for {\acsv} and $-0.0976$ for {\acsi}) corrects 10-pixel magnitude to infinite-aperture magnitude, and $\mathrm{ZP}_\mathrm{Vega}(\mathrm{MJD})$ is the time-dependent Vega magnitude zero point which ranges from 0.7335 to 0.7351 for {\acsv} and from 0.5294 to 0.5303 for {\acsi}. We present the fully-calibrated {\it HST} photometry measurements of stars used in this work in Table~\ref{tbl_mag}.

\begin{deluxetable}{ccccc}
\tabletypesize{\scriptsize}
\tablecaption{{\it HST} photometry\label{tbl_mag}
  for the 12 LMC fields}
\tablewidth{0pt}
\tablehead{
\colhead{Field} & \colhead{RA$^a$} & \colhead{Dec} & \multicolumn{2}{c}{Magnitudes$^b$}\\[-6pt]
\colhead{(NGC)} & \multicolumn{2}{c}{(J2000) [deg]} & \colhead{\acsi} &\colhead{\acsv}}
\startdata
1755 &    73.912541 &   -68.181942 &    17.098(3) &    18.245(4)\\
1755 &    73.927582 &   -68.190703 &    15.646(1) &    17.208(2)\\
1755 &    73.877665 &   -68.164196 &    16.857(3) &    18.145(4)\\
1755 &    73.888109 &   -68.170361 &    15.769(2) &    17.177(2)\\
1755 &    73.938231 &   -68.198904 &    16.959(3) &    16.928(2)\\
1755 &    73.926008 &   -68.195333 &    16.109(2) &    17.743(3)\\
1755 &    73.931482 &   -68.205091 &    17.233(3) &    18.239(4)\\
1755 &    73.867730 &   -68.170427 &    16.872(3) &    17.907(3)\\
1755 &    73.866620 &   -68.169905 &    16.594(2) &    17.612(3)\\
1755 &    73.869387 &   -68.172283 &    17.159(3) &    18.413(4)\\
\enddata
\tablecomments{$a$: Based on the OGLE catalog. $b$: Only list unsaturated sources with $F814W < 17.5$~mag. This table is available in its entirety in machine-readable form.}
\end{deluxetable}

\section{Methodology}\label{sec_met}

The ground-to-{\it HST} magnitude offset for a given star, $\Delta m \equiv F814W - I$, is dominated by filter response difference for objects in the outskirts of the LMC and by the combination of filter difference and blending for objects in crowded regions. Since the variation in crowding is relatively smooth across the LMC, we treat $\Delta m$ as a combination of these two factors. Statistically, the blending effect on the photometric measurement of a star is correlated with the local stellar number density and with its own brightness. In denser regions, there is a higher chance that background sources superpose onto the aperture where the photometry is measured; for a bright source, the superposed flux is relatively small compared to its own flux, leading to a reduced blending effect in magnitude space. The filter difference is common to all objects, and for stars it can be effectively modeled by a color term. We incorporate all these contributions by modeling the offset $\Delta m$ as a linear combination of three variables: the local stellar number density $N$ (defined below), the {\it HST} photometric measurement {\acsi}, and the ground color measurement $V\!-\!I$:
\begin{eqnarray} \label{equ_fit}
  \Delta m_I & = & a + b \cdot (N-17) + c \cdot (F814W-14.5)\nonumber\\
           &   & +\ d \cdot (V\!-\!I-1.6),
\end{eqnarray}
where $a$, $b$, $c$, and $d$ are free parameters. Dependent variable offsets were selected such that $a$ is the approximate magnitude offset for typical TRGB stars in the LMC.  We note that for such stars, the offset in extinction between the {\acsi} and $I$ filters varies only by $\sim \pm 3$~mmag for $0 < A_I < 0.2$ mag. Thus, we can safely treat this as a constant and use the term $a$ to absorb the overall difference.

The low resolving power of ground surveys can lead close sources to be merged into a single PSF, causing bias in the PSF modeling. A common practice to account for such bias is image level artificial star test\citep[e.g.][]{1988AJ.....96..909S,1996AJ....112.1928G,2006ApJS..166..534H,2015BaltA..24..314S} but usually limited to a small sample of stars given its intense computation cost. A good parametrization for this bias can be the luminosity density of those near-resolved sources, however, it is beyond the reach of this study where only catalog level data are available for both ground surveys. Instead, we adopted the local stellar number density to parameterize both the blending effect and possible crowding bias. For a fairly uniform initial mass function, measuring any star as a tracer is going to be fairly equivalent. We computed the local number density $N$ using the ``density sample'' as described in \S2 which consists of all the OGLE sources with $I<18$ mag. For each object, we adopted $N$ as the total number of stars in the density sample within a 20\arcsec\, radius around its location, which represents a fine enough local scale yet encircles a number count that is statistically significant. We tested other choices of the counting radius and found that smaller radii encircled too few stars that makes the result too noisy, while larger radii are not preferred because the bias is related to a local property. We compute $N$ using the OGLE density sample for both OGLE and MCPS analysis, as the OGLE catalog gives superior photometry thanks to its better seeing, finer pixel scale, and repeated observations.

We cross-matched the ground calibration samples to the {\it HST} measurements using matching radii of 0.26\arcsec\, for OGLE and 0.7\arcsec\, for MCPS, which are equivalent to the size of one pixel in their detectors. Due to the limited number of {\it HST} fields, the small field of view of ACS, and the limited range of overlapping magnitudes and colors, there were only 1208 OGLE and 1042 MCPS sources matched to the {\it HST} catalog with $14.5\,<\,I\,<\,17$~mag and  $0.5\,<\,V\!-\!I\,<\,2.5$~mag. Since each {\it HST} field contains a globular cluster, the local number density $N$ has a wide range of $3 < N < 40$, which extends far beyond the typical density of the LMC bar region ($10<N<30$).

\begin{figure}
\epsscale{1.2}
\plotone{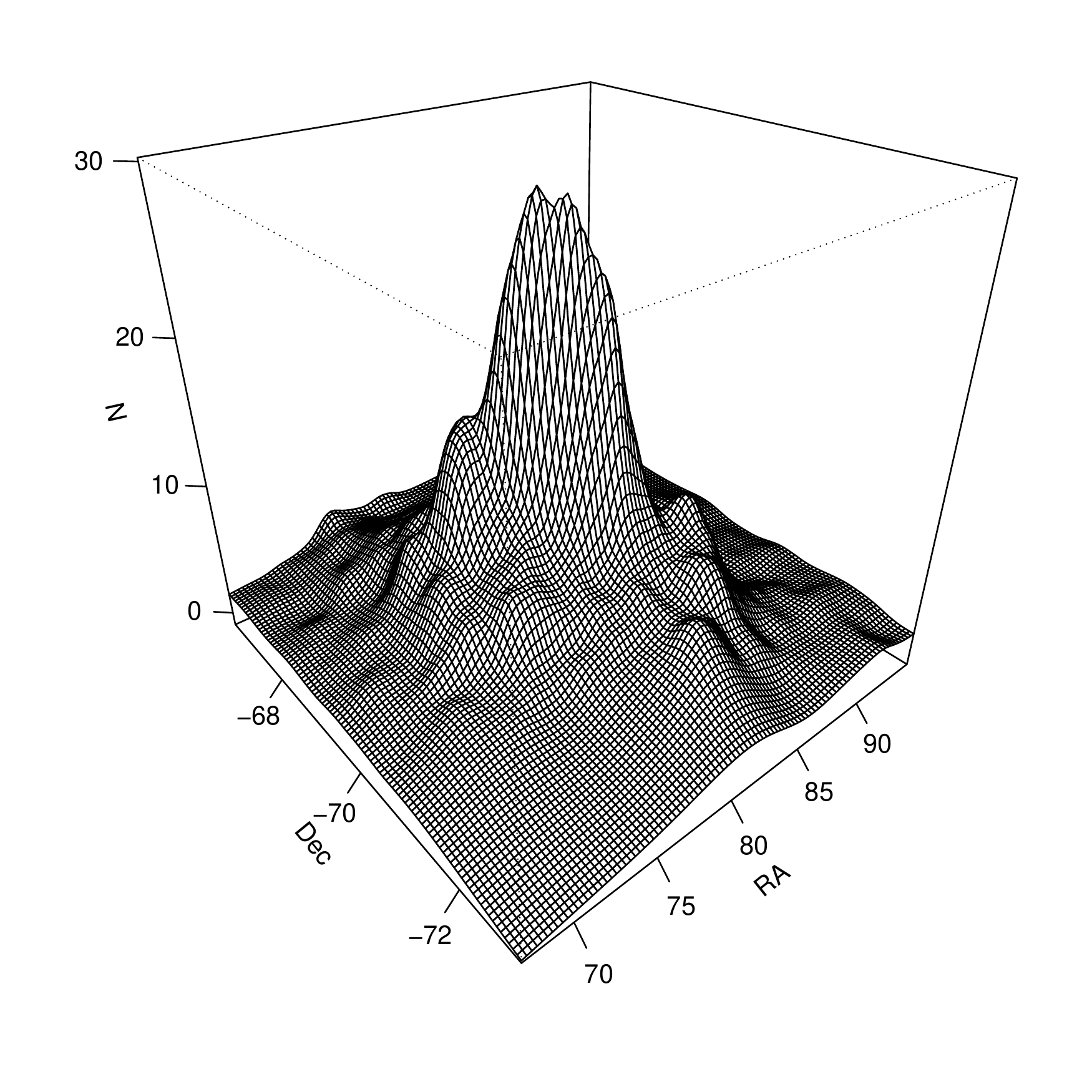}
\caption{Smoothed local number density distribution across the LMC obtained by a thin-plate spline fit. The bar region has a typical density of $10\lesssim N\lesssim 30$, where $N$ is the number of stars within 20\arcsec\, and $I<18$ mag. \label{fig_density}}
\end{figure}

\begin{figure*}
\epsscale{1.2}
\plotone{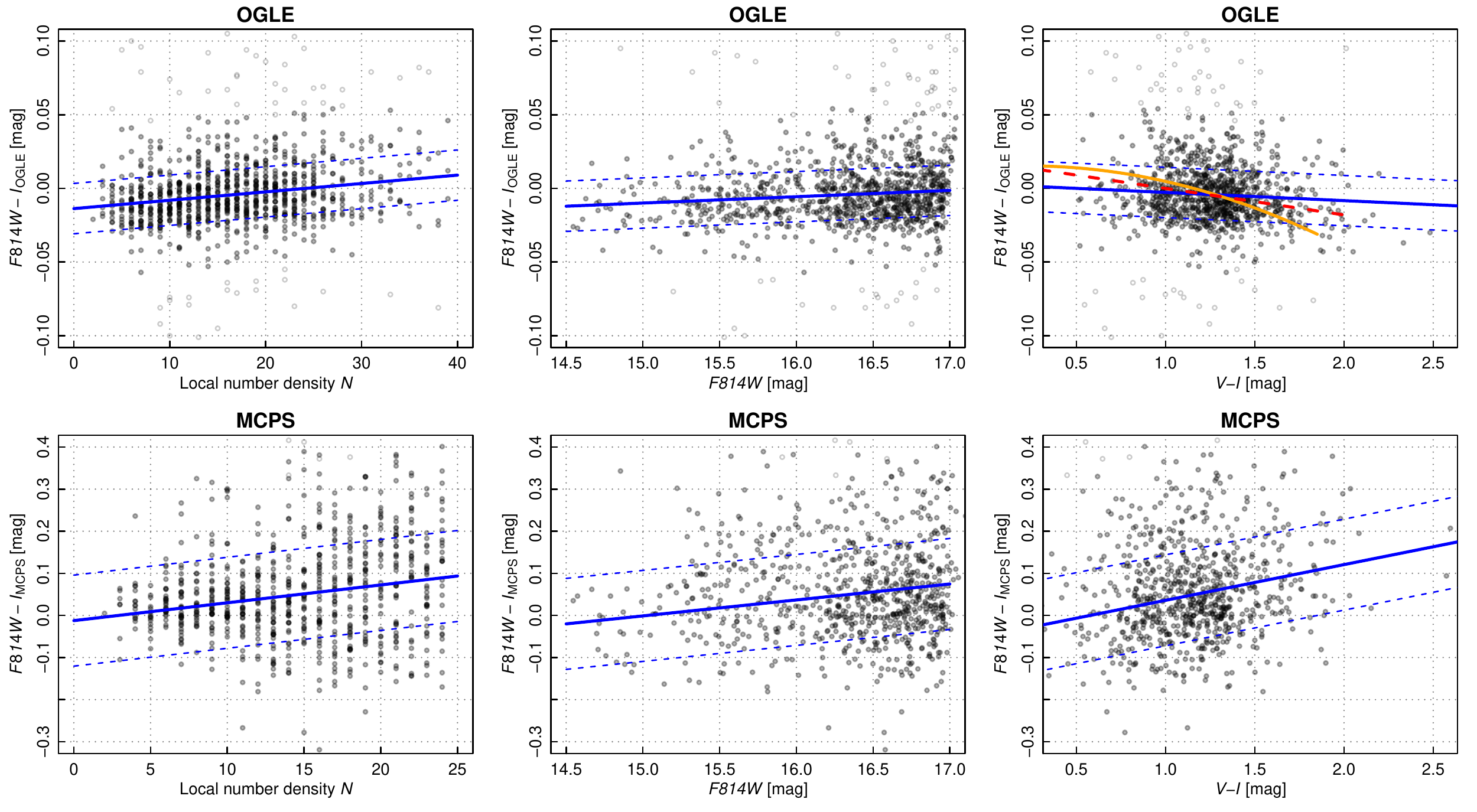}
\caption{Linear regression fit of Eqn.~\ref{equ_fit} to the OGLE (top) and MCPS (bottom) data in the overlapping {\it HST} fields within the LMC. Solid blue lines indicate the best-fit trends while dashed blue lines indicate the $\pm 1 \sigma$ dispersion of residuals. Open circles indicate rejected objects from iterative 3$\sigma$-clipping. The dashed red line shows the synthetic filter transformation from \citet{2016ApJ...826...56R} shifted by the WFC3-to-ACS offset. The solid orange line shows the quadratic synthetic transformation from \citet{2005PASP..117.1049S}. Note the different Y-axis ranges for the top and bottom  panels. The magnitude differences displayed in all panels are as measured, without shifting the mean offsets or detrending any dependent terms.\label{fig_fit}}
\end{figure*}

We fit Equation~\ref{equ_fit} to the matched sources with local number density restricted to $N<40$ for the OGLE sample and $N<25$ for the MCPS sample to avoid nonlinearity of the relations and large blending noise. Objects with local number densities beyond these limits should be excluded when deriving the TRGB offset using these calibrations, without losing much of the data (0.8\% for OGLE and 7.4\% for MCPS; the mean density distribution across the LMC is shown in Figure~\ref{fig_density}). We applied iterative $3\sigma$ clipping during the fit which rejected 8\% and 5\% of sources for OGLE and MCPS, respectively (reducing the outlier rejections to 1\% of sources changes the zeropoint of the relation by $<$ 3~mmag). The best-fit relations are shown in Figure~\ref{fig_fit}. Since we only use the bright end of the transformation for TRGB calibration, the small asymmetry in the residual density distribution at the faint end of the $\Delta m_I \sim$ {\it F814W} fit does not affect our results. We tested the fit with fainter stars rejected (cuts from 16 to 17 mag) and found the corresponding best-fit parameters change within the quoted uncertainties. We found the best-fit color term slope for $F814W - I_\textrm{OGLE}$ is shallower than the synthetic transformations from \citet{2005PASP..117.1049S} or for WFC3 from \citet{2016ApJ...826...56R}, but the mean offsets at the color of calibrating stars are comparable, as shown in Figure~\ref{fig_fit}. We also derived the transformations from ground $V$ magnitude to ACS {\acsv} magnitude
\begin{eqnarray} \label{equ_fitv}
  \Delta m_V &=& a + b \cdot (N-17) + c\cdot (F555W-15.8)\nonumber\\
             & & +\ d \cdot (V\!-\!I-1.6),
\end{eqnarray}
and summarized the coefficients in Table~\ref{tbl_coef}.

\begin{deluxetable}{lrrrrr}[b]
\tabletypesize{\scriptsize}
\tablecaption{Model coefficients for Equations~\ref{equ_fit} and \ref{equ_fitv}\label{tbl_coef}}
\tablewidth{0pt}
\tablehead{
\colhead{Band/}  & \colhead{$a^*$} & \colhead{$b^*$} & \colhead{$c^*$} & \colhead{$d^*$} & \colhead{scatter} \\[-6pt]
\colhead{Survey} &                 &                 &                &                 &\colhead{[mag]}}
\startdata
I/OGLE & -14(03) &     0.57(0.08) &      4.3(1.1) &     -5.6(2.2) &     0.017 \\
I/MCPS &  30(17) &      4.2(0.7) &   38(07) &   85(12) &     0.108\\
V/OGLE &  98(05) &     0.72(0.14) &     14.3(2.0) &     74.4(3.8) &     0.030 \\
V/MCPS & 162(21) &      7.2(0.8) &   57(09) &  -23(14) &     0.130
\enddata
\tablecomments{$a$: Offset. $b$: Stellar density. $c$: magnitude. $d$: $V\!-\!I$ color.\\ *: Values are multiplied by a factor of 1000.}
\end{deluxetable}

\begin{deluxetable}{lcccccccccccccccccccccccc}
\tabletypesize{\scriptsize}
\tablecaption{TRGB offsets between ground systems$^a$\label{tbl_off}}
\tablewidth{0pt}
\tablehead{
\colhead{Location}&&&&&&&&&&&&$\Delta V$ &&&&&&&&&&&&$\Delta I$ 
}
\startdata
LMC bar &&&&&&&&&&&& 0.12 &&&&&&&&&&&& 0.06 \\
LMC off-bar &&&&&&&&&&&& 0.04 &&&&&&&&&&&& 0.04 \\
SMC $r<30\arcmin$ &&&&&&&&&&&& 0.08 &&&&&&&&&&&& 0.04 \\
SMC $r<40\arcmin$ &&&&&&&&&&&& 0.07 &&&&&&&&&&&& 0.04
\enddata
\tablecomments{$a$: The offsets are calculated as OGLE$-$MCPS,
  reported in units of magnitude, and restricted to stars with $N<40$.}
\end{deluxetable}

\begin{figure*}[t]
\epsscale{1.2}
\plotone{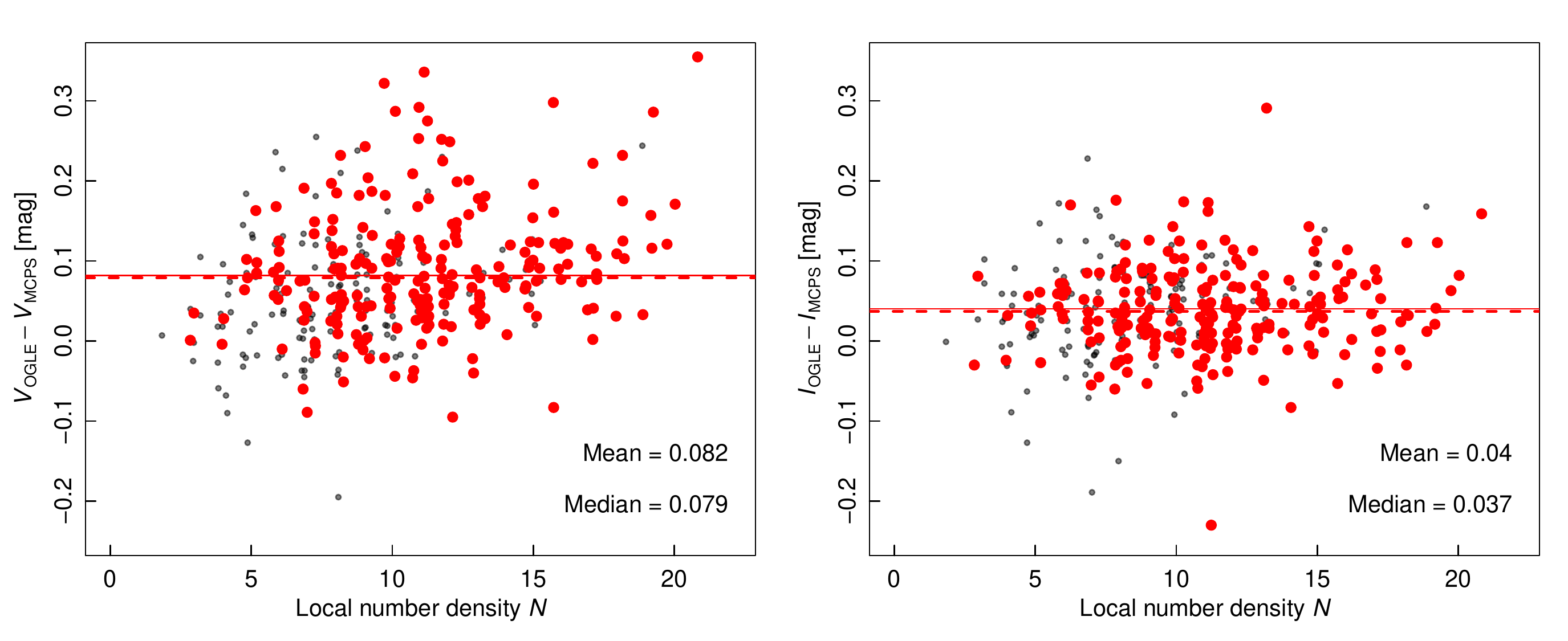}
\caption{$V$-band (left) and $I$-bang (right) magnitude differences between OGLE and MCPS for TRGB stars in two circular regions near the center of the SMC with radii of 30\arcmin\ (red), and 40\arcmin\ (red and black). The mean and median values for the red points are indicated by red solid lines and dashed lines, respectively, and displayed in the lower right of each panel.\\ \label{fig_smc}}
\end{figure*}

Independent of the comparison with {\it HST}, we also measured the difference in photometry for all TRGB-like stars (see criteria in \S 4.1) matched between the two ground-based catalogs in the LMC bar and off-bar regions (shown in Figure~\ref{fig_lmc}).  These differences are  summarized in Table~\ref{tbl_off}.  Although this data covers an area many orders of magnitude greater than the {\it HST} fields used to derive the transformations, the differences match the predictions of the fitting formula to better than 0.01~mag for TRGB-like stars with bar and off-bar densities of $N=25$ and $N=15$, respectively.  The differences between OGLE and MCPS are largest in the bar region, where the blending is highest, with mean OGLE$-$MCPS offsets of 0.12 and 0.06 mag for $V$ and $I$, respectively.  However, differences remain at the 0.04 mag level for both $V$ and $I$ off the bar, indicating the size of differences due to the photometric systems (zeropoints and color terms for red stars) even where stellar density decreases.

We also computed the magnitude differences between OGLE and MCPS measurements for the SMC as shown in Figure~\ref{fig_smc} and given in Table~\ref{tbl_off} (see \S 4.2 for details).  A similar difference to that of the LMC is also seen for the SMC, also increasing with local density.  Near the central region of the SMC, photometry from the two datasets differs by $\Delta V \sim 0.08$~mag and $\Delta I \sim 0.04$~mag. This has important consequences for the determination of the extinction towards the LMC obtained by F19, as discussed in \S 4.2.  

Considering the dramatic difference in blending effects, the OGLE data are vastly superior to the MCPS data for providing calibrations of TRGB in the Magellanic Clouds. In addition, the scatter of the MCPS fit is greater by a factor of 6.4 than the OGLE fit, likely due to the MCPS measurement uncertainties and blending noise. We conclude that the MCPS data is less favorable than the OGLE data for TRGB calibrations in the Magellanic Clouds and should not be used for this purpose.

\section{Results}

\subsection{Implications for the LMC TRGB}

The TRGB magnitude measures an ensemble property and usually requires a minimum of several hundred stars near its edge to reliably detect the change in the luminosity function. For the LMC, different regions (groups) can exhibit different metallicities and thus different loci of the Red Giant Branch \citep{2016MNRAS.455.1855C}. Therefore we studied the mean ground-to-{\it HST} transformations for different locations across the LMC.

\begin{figure*}
\epsscale{1.2}
\plotone{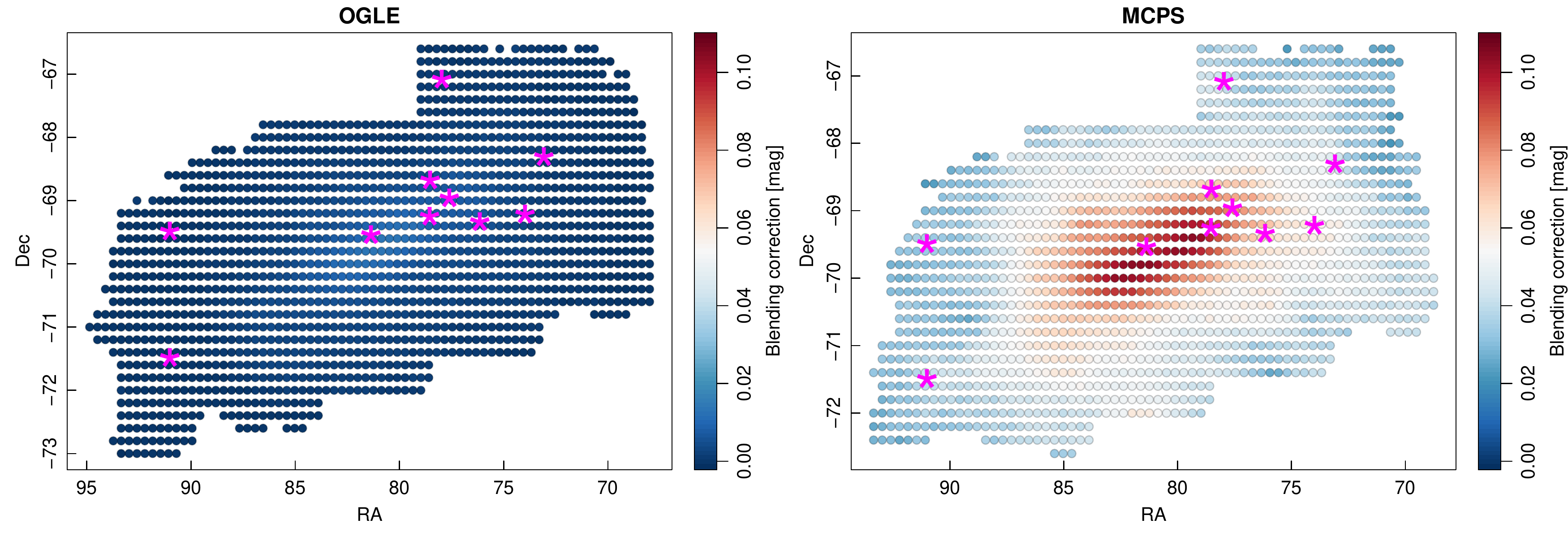}
\caption{Predicted mean blending corrections of TRGB stars across the LMC for the OGLE catalog (left) and MCPS catalog (right). Offsets due to filter transformation were subtracted for both panels. The magenta stars indicate the locations of ten fields studied by \citet{2017ApJ...835...28J}.\label{fig_ommap}}
\end{figure*}

We selected a complete sample of all LMC stars in the ground catalogs around the $I$-band TRGB magnitude from those matching the following criteria: (i) $14.5 < I < 14.7$ mag; (ii) $1.5 < (V\!-\!I) < 2.1$ mag; (iii) local number density $N < 40$. The resulting sample consists of 7088 objects spread across the galaxy. We computed the mean {\it HST}-to-ground offset, $\Delta m$, for the TRGB stars at a grid of positions with a group size of 20\arcmin. The mean blending corrections at different locations of the LMC are shown in Figure~\ref{fig_ommap}. As mentioned before, the MCPS data exhibit greater blending effect and are much less favorable for TRGB determinations. 

We calibrated the ACS {\acsi} TRGB in the LMC starting from the results of \citet[][hereafter, JL17]{2017ApJ...835...28J}, who measured the ground $I$-band TRGB magnitudes in ten fields of the LMC using the $VI$ photometry from the OGLE-III shallow survey. Because JL17 also measured the TRGB in SN Ia hosts with similar methods (i.e., smoothing and edge detection) as the LMC and utilized the data of the same OGLE system we analyzed, we chose not to independently measure the TRGB from the OGLE photometry but rather to apply our transformation to the JL17 values as listed in their Table~5 to the {\it HST} system. The JL17 values included corrections for the geometric separation of the regions from the LMC line of nodes and for extinction using the OGLE reddening maps of \citet{2011AJ....141..158H}. These maps, based on red clump stars and RR Lyrae, were shown in the original publication to be in strong agreement with maps derived from 2MASS data \citep{2008A&A...484..205D,2009AJ....137.5099D} and from the prior generation of OGLE-based maps \citep{2005A&A...430..421S}. Independently,  \citet{2018ApJ...860....1G} presents three estimates of extinction for 20 detached eclipsing binaries (DEBs; a younger population with more extinction than red giants) in the LMC, finding good agreement with the \citet{2011AJ....141..158H} maps with a mean $A_I$ that is 0.02~mag lower from measurements of Na I D1 lines and 0.03~mag higher from atmospheric modeling.  We will consider this subject further in the next section.

 We restricted the aforementioned TRGB sample where the {\it HST}-to-ground transformation was determined to $1.5 < (V\!-\!I) < 1.9$~mag and averaged the corrections of TRGB stars within 50\arcmin\, around each field in order to match the choices made by JL17. The mean corrections and {\acsi} TRGB magnitudes for the ten fields are summarized in Table~\ref{tbl_jl17}. The corrections for the OGLE system range from $-0.023$~mag in outer fields where the filter difference dominates, to $-0.012$~mag in crowded central regions where both the filter transformation (-0.023 mag) and the blending effect (+0.011 mag) contribute. The transformations for the MCPS system would be far greater and highly dependent on location but we discourage use of these as the MCPS photometry is too strongly contaminated by blending to be reliable to high precision.    

\begin{deluxetable}{cccc}
\tabletypesize{\scriptsize}
\tablecaption{LMC TRGB Magnitude in ACS {\it F814W}\label{tbl_jl17}}
\tablewidth{0pt}
\tablehead{
\colhead{Field} & \colhead{$I_0$ mag$^a$} & \colhead{$\Delta m^b$} & \colhead{{\it F814W}}
}
\startdata
EB1 &    14.525 $\pm$     0.035 &    -0.012 &    14.513\\
EB2 &    14.518 $\pm$     0.017 &    -0.013 &    14.505\\
EB3 &    14.531 $\pm$     0.025 &    -0.016 &    14.515\\
EB4 &    14.501 $\pm$     0.013 &    -0.022 &    14.479\\
EB5 &    14.476 $\pm$     0.023 &    -0.014 &    14.462\\
EB6 &    14.564 $\pm$     0.015 &    -0.016 &    14.548\\
EB7 &    14.565 $\pm$     0.016 &    -0.018 &    14.547\\
EB8 &    14.457 $\pm$     0.027 &    -0.020 &    14.437\\
CF1 &    14.483 $\pm$     0.025 &    -0.023 &    14.460\\
CF2 &    14.519 $\pm$     0.021 &    -0.022 &    14.497\\
\hline
Mean &    14.522 $\pm$     0.006 & $\dots$ &    14.504 $\pm$     0.006\\
Mean of CF1-2 &    14.504 $\pm$     0.016 & $\dots$ &    14.482 $\pm$     0.016\\
Mean of EB1-8 &    14.524 $\pm$     0.007 & $\dots$ &    14.507 $\pm$     0.007
\enddata
\tablecomments{$a$: Values and errors from Table~5 of \citet{2017ApJ...835...28J}. $b$: Combination of filter transformation and blending correction.}
\end{deluxetable}

\begin{figure}[b]
\epsscale{1.2}
\plotone{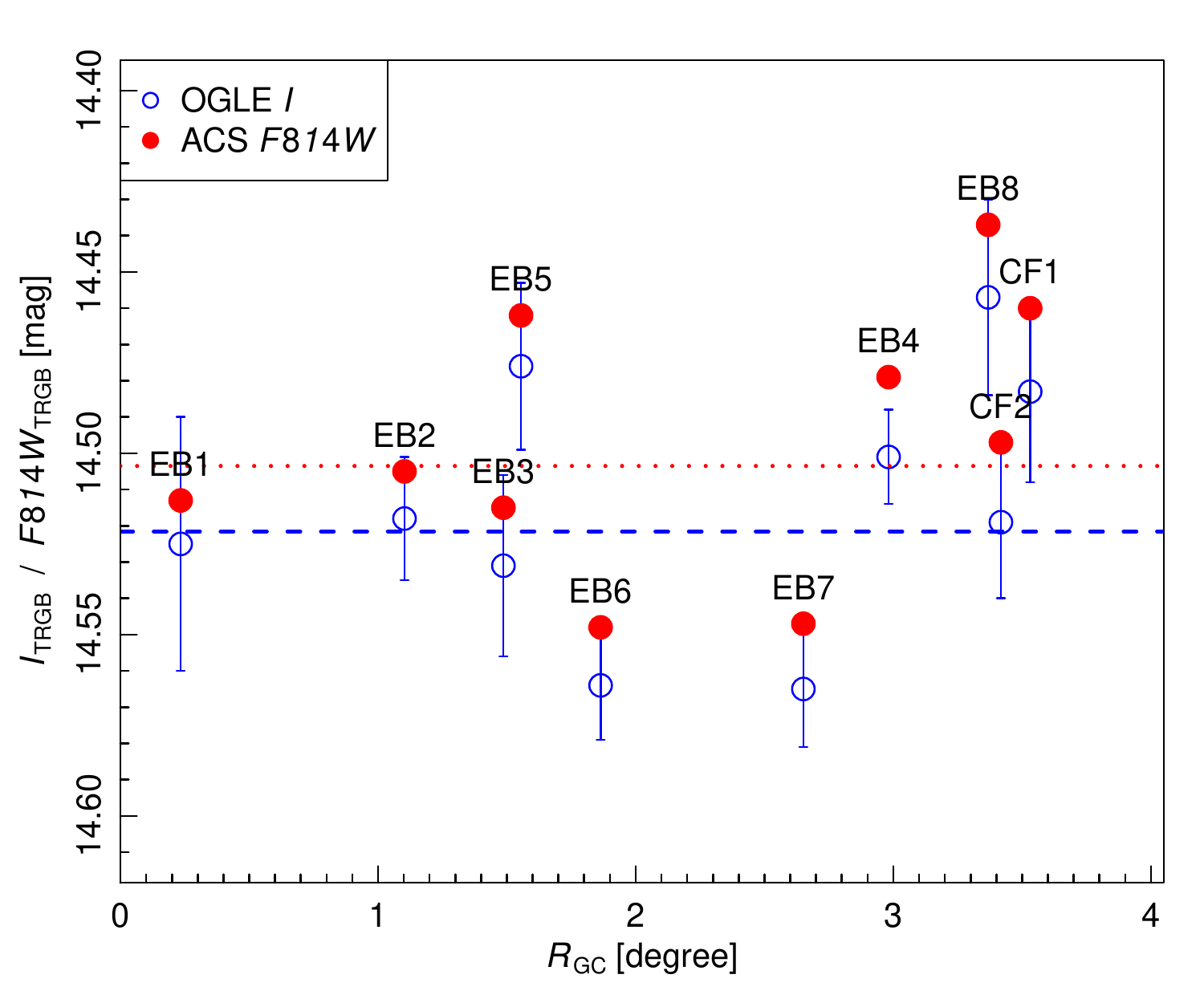}
\caption{Comparison of ground $I$ and ACS {\acsi} TRGB magnitudes (blue and red symbols, respectively) for the ten fields of JL17. The blue dashed and red dotted lines indicate the weighted means of the respective measurements.\label{fig_jl17}}
\end{figure}

The differences for ground $I$-band TRGB and ACS {\acsi} TRGB magnitudes are shown in Figure~\ref{fig_jl17}. By applying the empirical ground-to-{\it HST} correction we obtained a mean {\acsi} TRGB for the LMC of $14.507\pm 0.012 \textrm{(stat)}\pm 0.036 \textrm{(sys)}$~mag, where the quoted errors are: random error of 0.012~mag from the standard deviations of the eight EB fields scaled by the square root of the degrees of freedom; systematic error of 0.036~mag from a quadratic sum of extinction correction error (0.03 mag; Lee 2019, private communication\footnote{The uncertainties cited in \citet{2011AJ....141..158H} are based on 1$\sigma$ scatter of the colors for the selected red clump stars which are broadened by many factors beside differential reddening. The cited value here is based on the average scatter of the reddening in those ten fields studied by JL17.}) and intermediate-age population error (0.02 mag; JL17).  Using the geometric estimate of the distance to the LMC from \cite{2019Natur.567..200P} our best estimate of the absolute $I$-band LMC TRGB luminosity on the the ACS {\it HST} system is then $M_{F814W}=14.507-18.477=-3.97 \pm 0.046$ mag. Note this value includes a location-specific extinction correction from the OGLE extinction maps of \citet{2011AJ....141..158H} as part of the JL17 analysis.

\subsection{Impact of Blending on LMC TRGB\\Extinction Estimate and $H_0$}\label{sec_imp}

The TRGB is commonly observed in the halos of galaxies where extinction is a scant $A_I \sim 0.01$~mag, as indicated by the reddening of background quasars by foreground halos at radii from the host center of 10-20 kpc \citep{2010MNRAS.405.1025M}. However, there is strong motivation to calibrate the TRGB luminosity in the LMC where a precise $\sim 1\%$ geometric distance is now available \citep{2019Natur.567..200P}. However, the extinction of TRGB in the LMC is a much more significant $A_I \ge 0.1$ mag.  The line of sight extinction in optical bands outside the Milky Way can be difficult to estimate, and  uncertainties in this estimate can dominate the total uncertainty in the determination of the TRGB luminosity. 

\begin{figure}
\epsscale{1.2}
\plotone{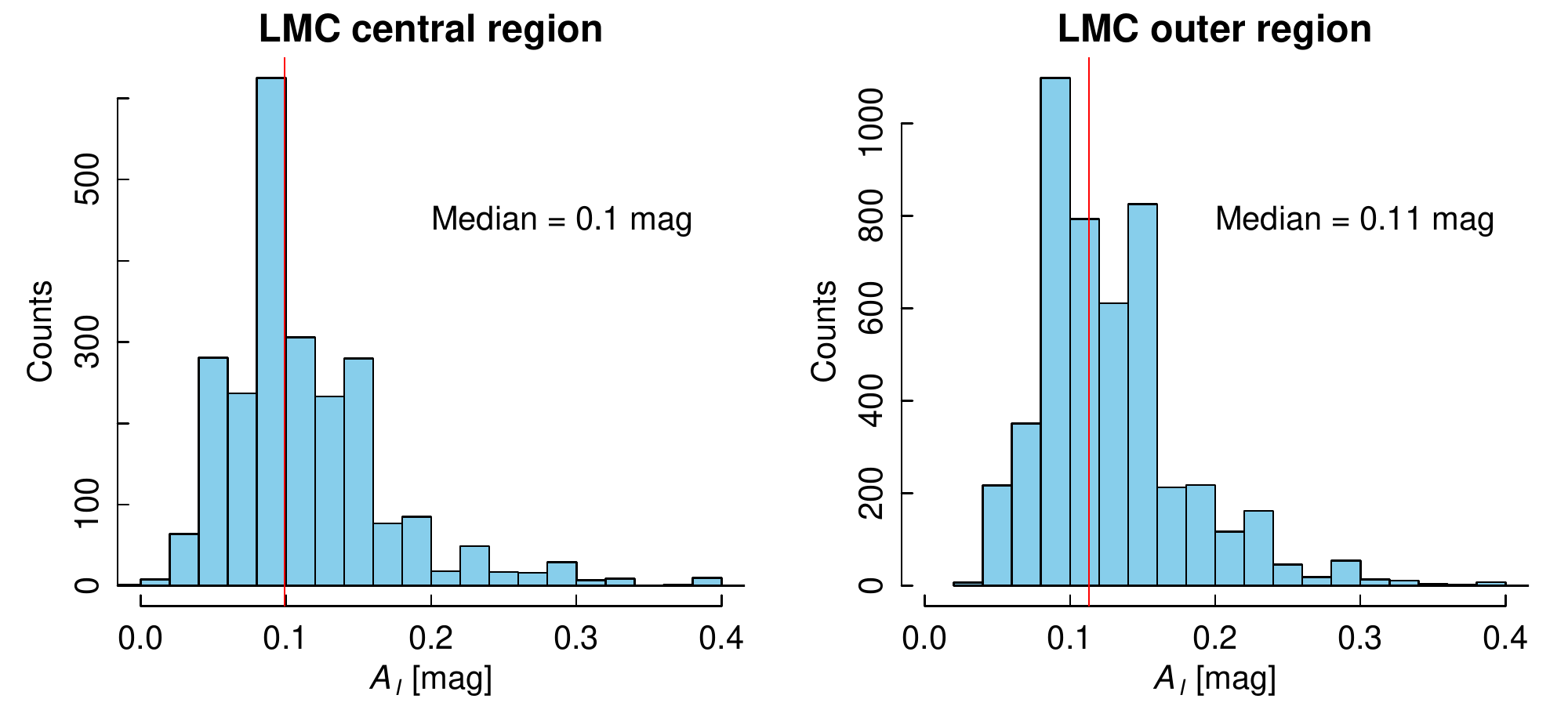}
\caption{$I$-band extinction for TRGB stars in the central (left) and outer (right) regions of the LMC based on \citet{2011AJ....141..158H}, which was adopted by JL17. The red vertical lines indicate the median values of $A_I$. \label{fig_lmcai}}
\end{figure}

 In the previous section we transformed the extinction-free, depth-corrected measurement of the LMC $I$-band TRGB from JL17 on the OGLE system to the {\it HST} ACS {\acsi} system based on matching photometry of sources on both systems.  JL17 used the extinction given at the locations of the TRGB stars in the OGLE reddening maps of \citet{2011AJ....141..158H} where location-specific line of sight extinction was measured from red clump stars and RR Lyrae.  Such stars are suitable tracers of interstellar extinction for TRGB since they come from similarly older or intermediate-age populations.
 The median LMC TRGB extinction given by these maps for all stars near the TRGB as identified in \S 4.1 is $A_I=0.10$~mag over the LMC bar region and $A_I=0.11$~mag in the region away from the bar (as shown in Figure~\ref{fig_lmcai}). Another recent and consistent estimate of the TRGB LMC extinction by \citet{2018ApJ...858...12H} found $A_I=0.06 \pm 0.06$~mag (from $E(B\!-\!V)=0.03 \pm 0.03$ mag) from a comparison of NIR TRGB colors across the LMC. 

A new estimate of the LMC TRGB extinction was made by F19 to calibrate TRGB and measure $H_0$ by comparing the LMC TRGB in the {\it VIJHK} bands to these same values derived in the SMC (and IC$\,$1613), where the TRGB extinction is estimated to be lower from its surrounding foreground extinction\citep{2011ApJ...737..103S}.  This comparison simultaneously constrains the parameters of relative distance and relative extinction, after which the extinction towards the SMC (or also IC$\,$1613) is  taken into account to yield the LMC value. F19 reported $A_I\textrm{(LMC)}=0.16 \pm 0.02$~mag, higher than the OGLE map value by $\sim 3 \sigma$.  

\begin{figure}
\epsscale{1.2}
\plotone{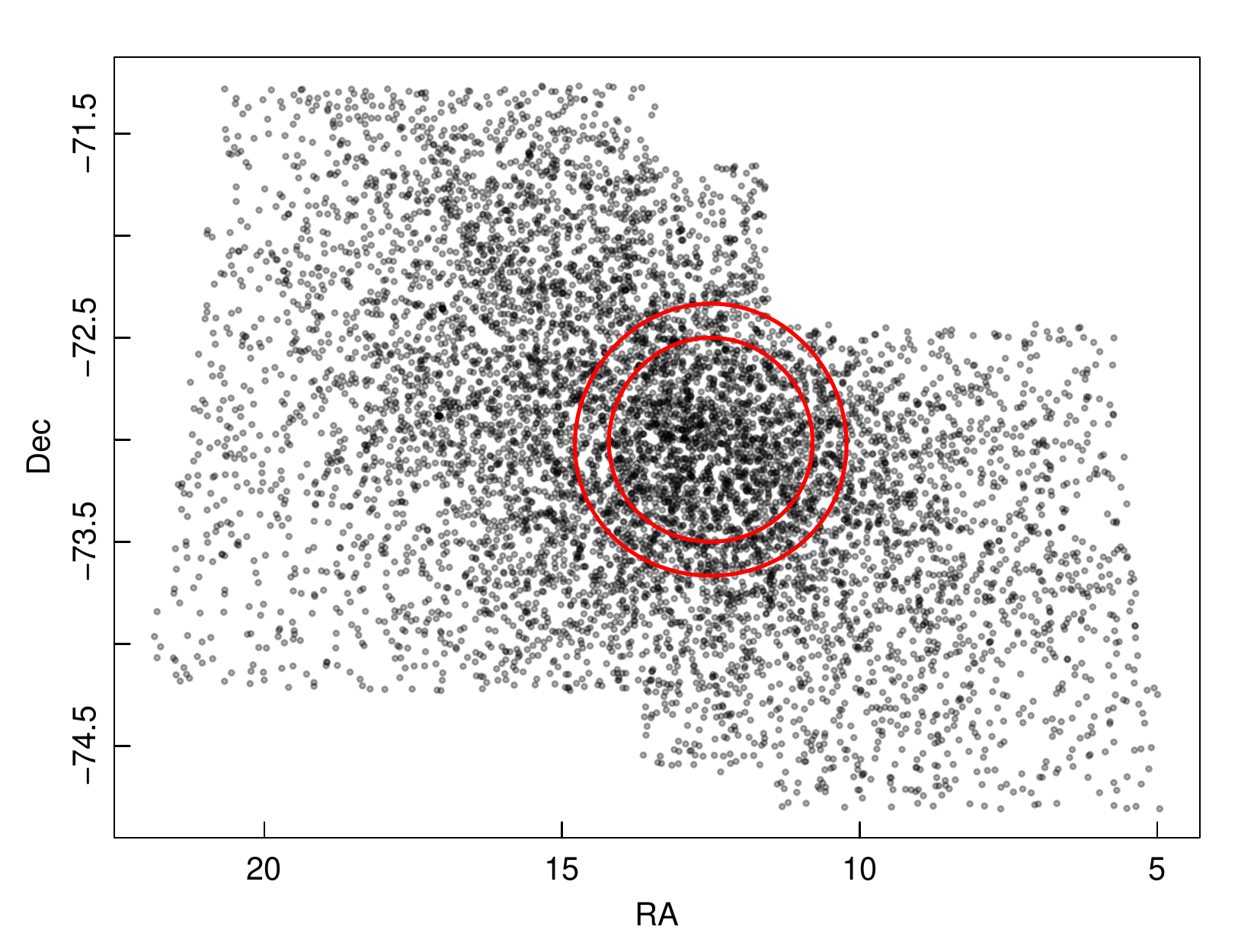}
\caption{Selected central (inner circle) and broader (outer circle) regions for magnitude comparison between OGLE and MCPS photometry of the SMC. \label{fig_smc2}}
\end{figure}

The comparison made by F19 of TRGB magnitudes between the Clouds utilized a different ground system for the photometry in the LMC than used for the SMC.  Specifically, F19 used OGLE-III photometry to measure the LMC TRGB and MCPS photometry \citep[from][]{2002AJ....123..855Z} for the SMC TRGB. Our direct comparison in \S3 of these two sources of ground photometry in the LMC revealed large offsets between them for TRGB stars that result from differences in blending, zeropoints and color terms.  The SMC TRGB photometry from MCPS used by F19 was obtained near the center of the galaxy and the LMC TRGB photometry was obtained away from the bar and assumes a mean matching the distance to the LMC line of nodes (Madore, private communication). Although the list of stars used by F19 has not been provided, Figure~\ref{fig_smc2} shows the positions of stars in the SMC derived from the OGLE catalog, from which we select two large circular regions centered on the galaxy (at $\alpha=12.5^\circ, \delta=-73^\circ$ in J2000) with radii of 30\arcmin\, and 40\arcmin. As previously discussed, Figure~\ref{fig_smc} showed a direct comparison of SMC OGLE and MCPS photometry in $V$ and $I$ for stars in these regions with magnitudes near the TRGB.  A significant inconsistency is apparent, similar to that seen in the LMC comparison.  In these central regions of the SMC, photometry given from the two datasets differs by $\Delta V \sim 0.08$ and $\Delta I \sim 0.04$~mag for TRGB stars in the inner 30\arcmin\, region and by $\Delta V \sim 0.07$ and $\Delta I \sim 0.04$~mag for stars within the broader 40\arcmin\, region (see Table~\ref{tbl_off}). To enable a check of this result, Table~\ref{tbl_smc} presents the coordinates and ground photometry in both systems for SMC TRGB stars in these regions. It is likely this difference arises from a similar combination of blending, zeropoint and color term differences as modeled in the LMC.  Regardless of its cause, this inconsistency between OGLE and MCPS photometry of TRGB-like stars in the SMC has important consequences for the  determination of the extinction towards the LMC obtained by F19.

\begin{deluxetable}{rrrrrr}
\tabletypesize{\scriptsize}
\tablecaption{SMC TRGB-like stars\label{tbl_smc}}
\tablewidth{0pt}
\tablehead{
\colhead{RA} & \colhead{Dec} & \multicolumn{2}{c}{OGLE [mag]} & \multicolumn{2}{c}{MCPS [mag]}\\
\multicolumn{2}{c}{(J2000) [deg]}& $V$ & $I$ & $V$ & $I$}
\startdata
   12.90837 &   -73.37339 &    16.390 &    14.918 &    16.351 &    14.884 \\
   13.11467 &   -73.40539 &    16.737 &    14.945 &    16.646 &    14.819 \\
   13.15875 &   -73.40647 &    16.531 &    14.957 &    16.450 &    14.795 \\
   13.45971 &   -73.43314 &    16.723 &    14.989 &    16.688 &    14.923 \\
   13.49960 &   -73.43797 &    16.724 &    15.025 &    16.488 &    14.900 \\
   13.50531 &   -73.38631 &    16.561 &    14.937 &    16.571 &    14.874 \\
   12.57692 &   -73.33008 &    16.640 &    14.906 &    16.548 &    14.874 \\
   12.79567 &   -73.35814 &    16.733 &    15.015 &    16.635 &    14.903 \\
   12.85271 &   -73.36336 &    16.544 &    14.982 &    16.366 &    14.923 \\
   12.99008 &   -73.35064 &    16.588 &    15.002 &    16.498 &    14.947 
\enddata
\tablecomments{This table is available in its entirety in machine-readable form.}
\end{deluxetable}

\begin{figure*}
\epsscale{1.2}
\plotone{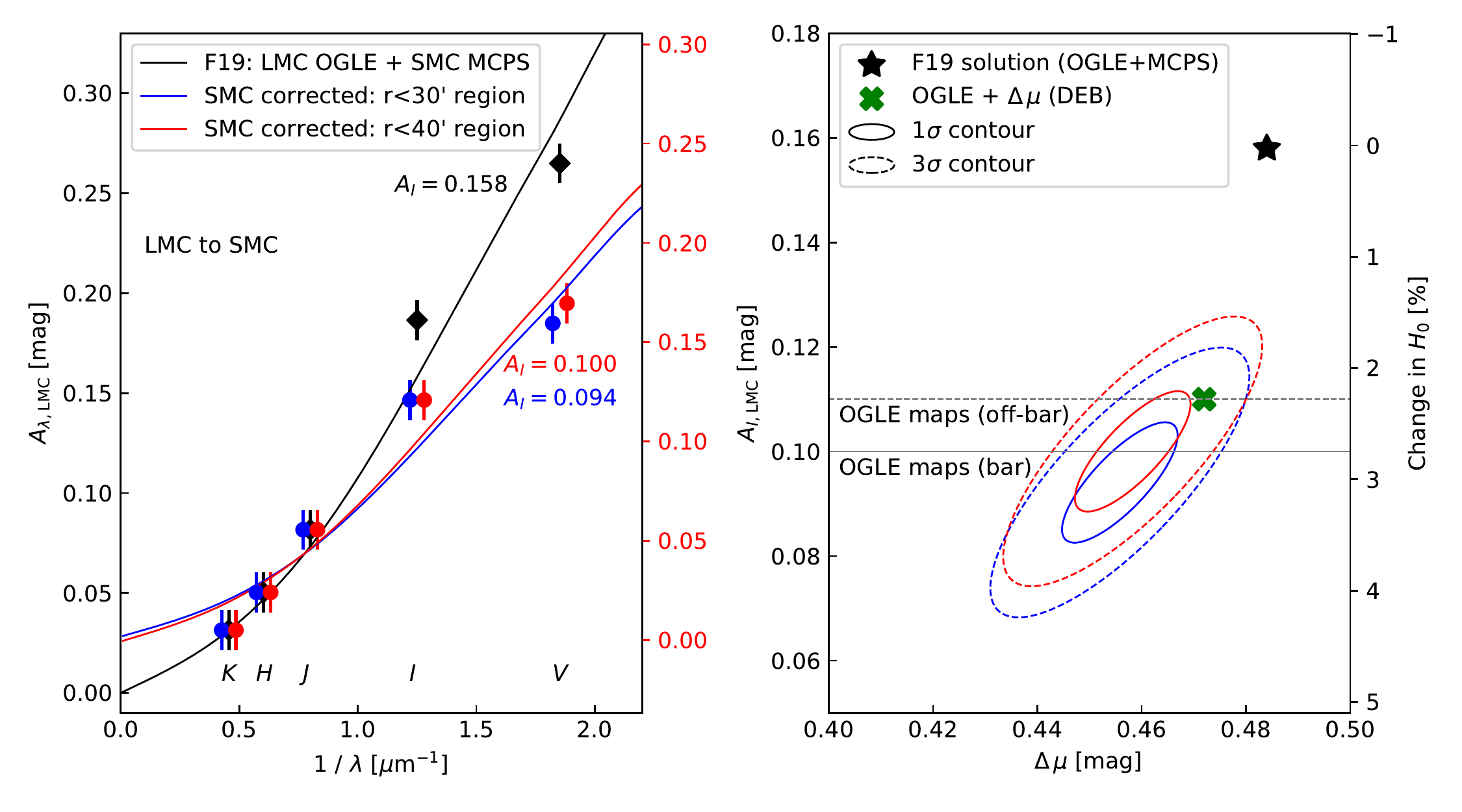}
\caption{Determination of LMC extinction based on LMC and SMC TRGB magnitudes at {\it VIJHK}, following the method of F19.  This method simultaneously constrains the relative distance modulus (a constant term) and the relative extinction (multiplier of reddening law).  The best-fit relative distance modulus between the Clouds is subtracted and the assumed SMC extinction is taken into account to arrive at the total LMC extinction. {\it Left}: Black points and line indicate the F19 result (their Fig.~A1). Blue and red lines show our two-parameter fits using corrected SMC photometry from the inner and broader regions (defined in the text), respectively. Points are slightly shifted in the horizontal direction to aid visualization. Note the different y-axis ranges for the black (left) and red/blue points (right), due to the difference in best-fit relative distance modulus. {\it Right}: Various solutions for the LMC $I$-band extinction and relative distance modulus between the Clouds, using the same color scheme as the left panel. Solid and dashed contours indicate $1$ and $3\sigma$ uncertainty ellipses. The green cross indicates the $I$-band extinction value obtained from OGLE TRGB photometry in both Clouds and the DEB relative distance moduli \citep{2017ApJ...842..116W}; it is not used in any of the fits. The median $I$-band extinction values for the bar \& off-bar regions of the LMC based on \citet{2011AJ....141..158H} are indicated by the horizontal black solid and dashed horizontal lines, respectively.\label{fig_rf19}}
\end{figure*}

We repeat the analysis presented in F19 by shifting {\it only} the SMC $V$ and $I$-band photometry to the OGLE system by the above amounts to account for the blending in the MCPS photometry relative to OGLE and so that the LMC and SMC measurements are based on the same OGLE system zeropoints and bandpasses. The results, adopting the \citet{1999PASP..111...63F} reddening law, are shown in Figure~\ref{fig_rf19}. Because the extinction parameter in the TRGB color comparison is more sensitive to the shorter wavelength $V$-band (the relative distance parameter is better constrained by the NIR data), the change in the estimated extinction is necessarily a large fraction of the change in $V$ photometry. The use of MCPS SMC photometry within the aforementioned broader and central regions leads to significant overestimates of the LMC extinction by $\Delta A_I$=0.06~mag. Accounting for this difference revises the estimate of the total LMC extinction by this method to $A_I=0.10$ and 0.09~mag, based on the broader and central SMC regions, respectively. This result is in good agreement with the median of $A_I=0.10$ and $A_I=0.11$~mag (for the LMC bar and outer regions, respectively) based on the \citet{2011AJ....141..158H} maps, as shown in Figure~\ref{fig_lmcai}. Our result is 0.06~mag or $3\sigma$ lower than the F19 estimate.

As a further check, we obtained an estimate of the LMC TRGB extinction through a comparison to the SMC independent of the analysis in F19. We measured the $I$-band TRGB for both Clouds using only OGLE data.  We removed the expected relative blending of the photometry between the Clouds indicated by our fitting formulae ($\sim$0.01 mag) and corrected for the geometric inclination of the LMC using the geometry solved by \citet{2019Natur.567..200P}. We obtained $I$-band TRGB magnitudes of 14.61 and 15.01 for the LMC and SMC, respectively. Since no extinction correction has been applied to either measurement, the difference between these two values ($0.40\pm0.02$~mag) is a combination of relative distance modulus and relative $I$-band extinction between the Clouds. That is, $\Delta \textrm{mag}=\Delta \mu+A_I\mathrm{(SMC)}-A_I\mathrm{(LMC)}$. \citet{2017ApJ...842..116W} provides the most precise relative distance modulus to date, $\Delta \mu = 0.472 \pm 0.026$~mag, by using DEBs systems in both Clouds, a robust approach thanks to the cancellation of calibration systematics. The advantages of this route to measure the LMC extinction is that it only makes use of TRGB measurements in $I$, which is the only band where this distance indicator is known to be quite insensitive to differences in star formation history (SFH) and metallicity.  By subtracting the aforementioned relative distance modulus we obtain a relative $I$-band extinction of 0.07 mag between the Clouds. Assuming the same SMC extinction adopted by F19 of $A_I\mathrm{(SMC)}$=0.037~mag, the total extinction towards the LMC is then $A_I\mathrm{(LMC)}=0.11\pm 0.03$~mag (see Figure~\ref{fig_lmcai}), consistent with our previous analysis and the average extinction from \citet{2011AJ....141..158H}. We note that the relative TRGB magnitude changes marginally if adopt the quadratic color-corrected $QT$ magnitude introduced by JL17, and in the direction of lower extinction towards the LMC if we adopt the linear color-corrected $T$ magnitude introduced by \citet{2009ApJ...690..389M}.

\subsection{TRGB Dependence on Metallicity and SFH}

An additional extinction estimate for the LMC is provided by F19 through a comparison of {\it VIJHK} TRGB colors between the LMC and IC$\,$1613. The TRGB IC$\,$1613 detections are presented in \citet{2017ApJ...845..146H} where ground-based $V$-band photometry is calibrated to a redder ACS filter, {\acsr}.  Due to the large color term which results from the comparison, \citet{2017ApJ...845..146H} state that this $V$-band data is useful only to identify tip sources and not suitable for the actual TRGB measurement. As a result, and because our transformations were derived between $V$ and {\acsv}, we are unable to reanalyze the IC$\,$1613 data on a matching photometric systems as was done between the LMC and SMC.

\begin{figure*}
\epsscale{1.2}
\plotone{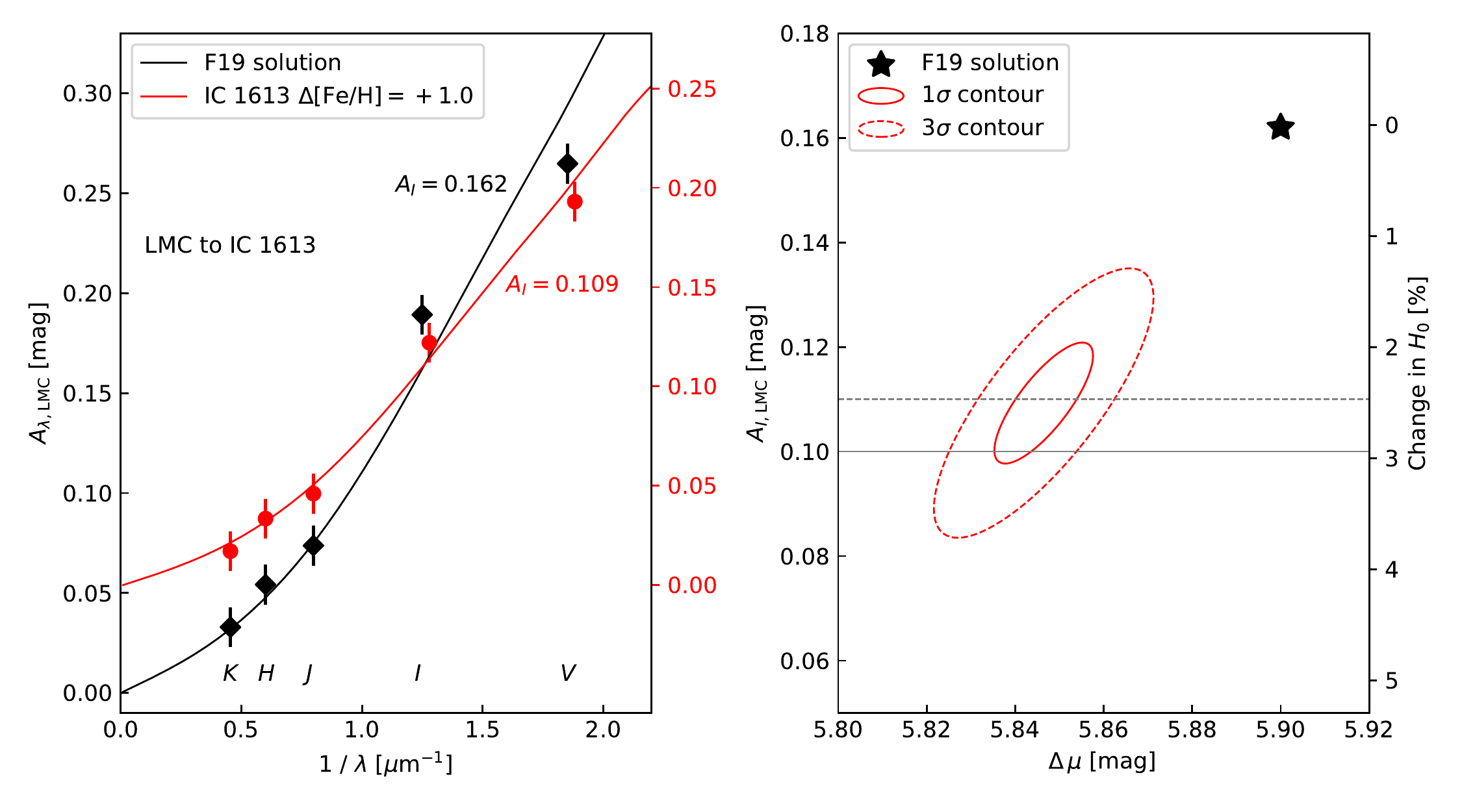}
\caption{Same as Fig.~\ref{fig_rf19} but for a comparison of TRGB magnitudes in the LMC and IC$\,$1613, following the method of F19 (see their Fig.~A2). The TRGB magnitudes for IC$\,$1613 \citep{2017ApJ...845..146H} have been corrected for $\Delta$[Fe/H]$=1$~dex following \citet[][see text for details]{2019arXiv190401571M}.  This reduces the inferred LMC extinction to $A_I = 0.11$~mag, in better agreement with the OGLE reddening maps.  We consider this impact of metallicity illustrative of the size and scale of the impact on inferred extinction but would require greater understanding to be reliable. \label{fig_ic}}
\end{figure*}

However, the F19 comparison between LMC and IC$\,$1613 TRGB colors introduces an additional complication worth considering, given the sensitivity of the location of the TRGB in the CMD as a function of metallicity and SFH. The metallicity of the TRGB in IC$\,$1613 is [Fe/H] =$-1.6\pm0.2$~dex, compared to $\sim$ $-0.5$~dex for the LMC and $-1$~dex for the SMC \citep{2017ApJ...834...78M, 2019arXiv190103448N}. According to \citet{2019arXiv190401571M}, the use of a single slope to ``rectify'' the color dependence of TRGB at different metallicities produces differences in the TRGB of $\sim0.04$~mag for $\Delta$[Fe/H]=1~dex at the same SFH (from -2 to -1~dex; see their Fig.~7).

F19 assign total uncertainties of 0.007 mag for the rectified magnitudes of TRGB in each band of each host, 0.01 mag as their quadrature sum when comparing two hosts (see their Figures~A1 and A2).   The modeled $\sim0.04$~mag differences due to metallicity are far greater than these uncertainties and more importantly, the modeling indicates they have a systematic shift with wavelength.  To illustrate the impact of this on the extinction estimate, we naively use the comparison in \citet{2019arXiv190401571M} for the same SFH and $\Delta$[Fe/H]=1 (from -2 to -1~dex). This indicates that the rectified TRGB colors of the metal-poorer host, in this case IC$\,$1613, would be fainter in $J,H$ and $K$ by $\sim$ 0.026, 0.033 and 0.038~mag, respectively, and brighter in $V$ and $I$ by $\sim$ 0.019 and 0.014~mag, respectively. This wavelength dependence of the metallicity effect reduces the implied extinction of the metal-rich host (in this case the LMC) relative to the metal-poor one and yields a total LMC extinction of $A_I$=0.10~mag, in agreement with the earlier determinations. Figure~\ref{fig_ic} shows the impact of these metallicity corrections on the derived LMC extinction. Even without this systematic change with wavelength, allowing for a random error of the modeled size of 0.04 mag per band due to comparing TRGB across hosts with a metallicity difference of $\sim$ 1.0 dex and in bands other than the $I$-band would increase the uncertainty in the LMC extinction to 0.08 mag, reducing the value of this comparison.

However, we caution that a correction to TRGB colors may depend on both SFH and metallicity, which may require additional modeling and likely incurs uncertainties that may make the resulting estimate unreliable. Additional evidence of greater complexity in TRGB in the NIR comes from \citet{2012ApJS..198....6D}, who compared the TRGB in $J$ and $H$ across 23 nearby galaxies using {\it HST} imaging and found a scatter of 0.05~mag and offsets between these and globular cluster TRGBs and model TRGBs at the 0.1~mag level. Here we take only {\it the direction and scale} from comparing metal-poor to more metal-rich hosts as suggestive that this comparison is challenging and likely too uncertain at present to support the small extinction uncertainty in F19 of 0.02 mag and a revision of the OGLE map-based extinction estimate.  

Finally, we note that the quality of the fit to the reddening between the LMC and the other hosts with the adopted F19 relative errors of 0.01~mag is quite poor, with $\chi^2=13.7$ and 16.4 for 3 degrees of freedom (5 data points minus 2 free parameters; probabilities to exceed these $\chi^2$ are 0.3\% and 0.1\%) for the SMC and IC 1613 comparison, respectively.  This suggests that there may indeed be additional variations due to SFH and metallicity that are well in excess of the estimated errors and that this approach may require additional understanding before its robustness can be established, even for the SMC corrected version of the LMC--SMC comparison (Figure~\ref{fig_rf19}).

From the above we conclude that the LMC TRGB extinction estimates from \citet{2011AJ....141..158H} and employed by JL17 appear reasonable and our best estimate of the LMC TRGB luminosity on the ACS {\it HST} system is then $M_{F814W}=-3.97 \pm 0.046$ mag. The difference between this value and the one in F19 is 0.079~mag or 3.7\% in $H_0$, increasing the F19 estimate of $H_0$ to 72.4 $\pm 2.0$ km s$^{-1}$ Mpc$^{-1}$. Most of this difference, 0.06~mag, results from the described differences in LMC extinction estimators.

\section{Summary}

In this work, we addressed the cross-instrument zero point issue for the LMC TRGB determination by analyzing two ground datasets (OGLE and MCPS) and {\it HST} observations. We studied the blending effect and filter transformations for these ground data by comparing their photometry directly to ACS {\acsi} observations of the same fields, with the latter camera and filter being widely used for TRGB measurements in extragalactic systems for the cosmic distance ladder.

We found that the MCPS data, compared to OGLE, are less suitable for precision TRGB studies in both Magellanic Clouds due to severe biases caused by blending. In the bar region of the LMC, where most TRGB stars are located, the blending effect biased the MCPS photometry by $\sim$ 0.06--0.1~mag, which is greater than the current precision of $H_0$ measurement of 0.04 mag. On the other hand, the OGLE data exhibit a much reduced and correctable blending bias ($\sim$ 0.01 mag) for most sources, making them a better choice for TRGB calibrations.

The filter transformation between OGLE $I$ and ACS {\acsi} is relatively small but non-negligible. For the magnitude and color ranges of TRGB stars in the LMC, the offset solely due to filter transformation is $-0.023$ mag. This offset is empirically determined for the the OGLE data, compared to a value of $-0.011$~mag adopted by F19. For OGLE $V$ and ACS {\acsv} the the offset is strongly color-dependent, as the ACS {\acsv} filter response significantly extends towards the bluer direction. Here we report an empirically determined color transformation between OGLE $V\!-\!I$ and ACS {\acsv}$-${\acsi} based on 929 stars in the LMC with $0.5 < V\!-\!I< 2.3$
\begin{equation*}
  F555W-F814W = 1.082 ({\pm 0.001}) \cdot(V\!-\!I),
\end{equation*}
which may be useful for future color-dependent calibrations of the TRGB.

We derived ground-to-{\it HST} correction formulae to account for the blending effect and filter transformations in the LMC that explain the large differences seen between two commonly used ground systems.   Because these observed differences are a particularly large $\sim 0.1$~mag for MCPS TRGB photometry of the SMC in the $V$-band which was utilized in a comparison by F19 to determine the LMC extinction we find this inconsistency between ground systems leads to an overestimate of the LMC extinction by $\sim 0.06$~mag.  The extinction estimate of $A_I\mathrm{(LMC)}$=0.10~mag resulting from correcting the MCPS SMC photometry for blending was confirmed with our own estimate of the TRGB in the LMC and SMC using a consistent set of $I$-band photometry from OGLE and the relative distance between the Clouds from DEBs.  This value further matches the mean value from the OGLE reddening maps derived from red clump stars at the positions of the TRGB stars in the LMC from \citet{2011AJ....141..158H} and employed by JL17 for their TRGB calibration.  We therefore started with the JL17 TRGB calibration and transformed it to the ACS {\acsi} system, obtaining $M(\textit{F814W})=-3.97\pm 0.046$~mag and a TRGB-based $H_0$ which is 3.7\% higher than the F19 value. Though the ground-to-{\it HST} filter transformation and blending effect of the OGLE data in the LMC is not negligible, it is the re-evaluation of the LMC extinction from a more consistent dataset for the SMC that explains most of the $H_0$ discrepancy between F19 and this study.

While the {\it application} of TRGB in the halos of SN Ia hosts is quite insensitive to extinction, its calibration in the LMC is highly dependent on the estimate of its considerable extinction.  Additional studies of the extinction of TRGB stars in the LMC is warranted as well as the use of additional geometric distance anchors.  In the future, parallax estimates from Gaia out of the plane (and dust) of the Milky Way to TRGB stars and improvements in the distance estimates from masers in NGC 4258 are likely to alleviate the present reliance on the LMC and estimates of its extinction.

\acknowledgments
{We thank Kristin McQuinn for useful discussions about TRGB and metallicity.  We thank Dennis Zaritsky for helpful discussions concerning the MCPS Survey.  We also thank In Sung Jang, Barry Madore, Wendy Freedman, Myung Gyoon Lee, Rachael Beaton, Lars Bildsten and Brent Tully for useful discussions concerning application of the TRGB method.

Support for this work was provided by the National Aeronautics and Space Administration (NASA) through programs GO-14648 and GO-15146 from the Space Telescope Science Institute (STScI). Some of the data presented in this paper were obtained from the Multimission Archive at STScI (MAST). STScI is operated by the Association of Universities for Research in Astronomy, Inc., under NASA contract NAS5-26555.}

\facilities{HST (ACS), MAST}

\bibliographystyle{aasjournal}
\bibliography{lmct}

\end{document}